%Paper: hep-th/9403058
%From: James H. Horne <jhh@genesis5.physics.yale.edu>
%Date: Wed, 9 Mar 94 18:12:31 -0500

\input harvmac

% Four postscript figures accompany this paper. They are in a second part
% as a uuencoded compressed tar file with instructions for unpacking. If
% you can't handle figures, comment out the following line.
\input epsf

% Macro plagiarized from P.G.

\def\figin{\epsfcheck\figin}\def\figins{\epsfcheck\figins}
\def\epsfcheck{\ifx\epsfbox\UnDeFiNeD
\message{(NO epsf.tex, FIGURES WILL BE IGNORED)}
\gdef\figin##1{\vskip2in}\gdef\figins##1{\hskip.5in}% blank space instead
\else\message{(FIGURES WILL BE INCLUDED)}%
\gdef\figin##1{##1}\gdef\figins##1{##1}\fi}
\def\DefWarn#1{}
\def\figinsert{\goodbreak\midinsert}
\def\ifig#1#2#3{\DefWarn#1\xdef#1{fig.~\the\figno}
\writedef{#1\leftbracket fig.\noexpand~\the\figno}%
\figinsert\figin{\centerline{#3}}\medskip\centerline{\vbox{\baselineskip12pt
\advance\hsize by -1truein\noindent\footnotefont{\bf Fig.~\the\figno:} #2}}
\bigskip\endinsert\global\advance\figno by1}

\def\inbar{\,\vrule height1.5ex width.4pt depth0pt}
\font\cmss=cmss10 \font\cmsss=cmss10 at 7pt
\def\IR{\relax{\rm I\kern-.18em R}}
\def\IZ{\relax\ifmmode\mathchoice
{\hbox{\cmss Z\kern-.4em Z}}{\hbox{\cmss Z\kern-.4em Z}}
{\lower.9pt\hbox{\cmsss Z\kern-.4em Z}}
{\lower1.2pt\hbox{\cmsss Z\kern-.4em Z}}\else{\cmss Z\kern-.4em Z}\fi}
\def\IC{\relax\hbox{$\inbar\kern-.3em{\rm C}$}}
\def\IGa{\relax\hbox{${\rm I}\kern-.18em\Gamma$}}
\def\IPi{\relax\hbox{${\rm I}\kern-.18em\Pi$}}
\def\ITh{\relax\hbox{$\inbar\kern-.3em\Theta$}}
\def\IOm{\relax\hbox{$\inbar\kern-3.00pt\Omega$}}
\def\CM {{\cal M}}
\def\CF {{\cal F}}

\def\CL {{\cal L}}

\def\p {\partial}
\def\CS {{\cal S}}

\def\half{{1 \over 2}}

\def\Im{\mathop{\rm Im}}
\def\Re{\mathop{\rm Re}}
\def\Tr{\mathop{\rm Tr}}
\def\deg{\mathop{\rm deg}}
\def\mod{\mathop{\rm mod}}

%\noblackbox

\lref\banks{T.~Banks, D.B.~Kaplan, and A.E.~Nelson, ``Cosmological
Implications of Dynamical Supersymmetry Breaking,'' Aug.~1993,
hep-ph/9308292.}
\lref\brustein{R.~Brustein and P.J.~Steinhardt,
``Challenges for Superstring Cosmology,''
{\it Phys. Lett.} {\bf B302} (1993) 196, hep-th/9212049.}
\lref\carlos{B.~de Carlos, J.A.~Casas, and C.~Munoz,
``Supersymmetry Breaking and Determination of the Unification Gauge
Coupling Constant in String Theories,''
{\it Nucl. Phys.} {\bf B399} (1993) 623, hep-th/9204012.}
\lref\cjsf{E.~Cremmer, B.~Julia, J.~Scherk, S.~Ferrara, L.~Girardello,
and P.~van Nieuwenhuizen, {\it Nucl. Phys.} {\bf B147} (1979) 105;
E.~Cremmer, S.~Ferrara, L.~Girardello, and A.~Van Proeyen, {\it
Nucl. Phys.} {\bf B212} (1983) 413.}
\lref\dine{M. Dine and N. Seiberg,
``Is the Superstring Weakly Coupled?''
{\it Phys. Lett.} {\bf B162} (1985) 299.}
\lref\dineb{M.~Dine and N.~Seiberg,
``Is the Superstring Semiclassical?''
in {\it Unified String Theories}, ed.~M.B.~Green and D.J.~Gross (World
Scientific, Singapore, 1986) 678.}
\lref\giveon{A. Giveon, M. Porrati, and E. Rabinovici, ``Target Space
Duality in String Theory,'' Jan.~1994, hep-th/9401139.}
\lref\langacker{P.~Langacker and N.~Polonsky,
``Uncertainties in Coupling Constant Unification,''
{\it Phys. Rev.} {\bf D47} (1993) 4028, hep-ph/9210235.}
\lref\lerche{A.N.~Schellekens and N.P.~Warner,
``Anomalies, Characters and Strings,''
{\it Nucl. Phys.} {\bf B287} (1987) 317; W.~Lerche, B.E.W.~Nilsson,
A.N.~Schellekens, and N.P.~Warner,
``Anomaly Cancelling Terms from the Elliptic Genus,''
{\it Nucl. Phys.} {\bf B299} (1988) 91.}
\lref\mayr{P.~Mayr, H.P.~Nilles, and S.~Stieberger,
``String Unification and Threshold Corrections,''
{\it Phys. Lett.} {\bf B317} (1993) 53, hep-th/9307171.}
\lref\finite{G. Moore, ``Finite in All Directions,'' May 1993,
hep-th/9305139.}
\lref\serre{J.-P. Serre, {\it A Course in Arithmetic} (Springer-Verlag,
New York, 1973).}
\lref\siegel{W.~Seigel, ``Manifest Duality in Low-Energy Superstrings,''
Aug.~1993, hep-th/9308133.}
\lref\mautner{F.I. Mautner, ``Geodesic Flows on Symmetric Riemann
Spaces,''  {\it Ann. Math.} {\bf 65} (1957) 416.}
\lref\ccmr{C.C. Moore, ``Ergodicity of Flows on Homogeneous
Spaces,'' {\it Amer. J. Math.} {\bf 88} (1966) 154.}

\Title{\vbox{\baselineskip12pt\hbox{YCTP-P2-94}
\hbox{RU-94-25}
\hbox{hep-th/9403058}
}}
{\vbox{\centerline{Chaotic Coupling Constants}
}}

\centerline{{James H. Horne}\footnote{$^\dagger$}{Email address:
jhh@waldzell.physics.yale.edu} and Gregory
Moore\footnote{$^*$}{On leave from Yale. Email address:
moore@castalia.physics.yale.edu} }
\vskip.12in
\centerline{\sl $^{\dagger,*}$Department of Physics}
\centerline{\sl Yale University}
\centerline{\sl New Haven, CT 06520-8120}
\vskip .12in
\centerline{\sl $^{*}$ Department of Physics}
\centerline{\sl Rutgers University}
\centerline{\sl Piscataway, NJ, 08855}

\bigskip
\centerline{\bf Abstract}

We examine some novel physical consequences of the general structure
of moduli spaces of string vacua.  These include (1) finiteness of the
volume of the moduli space and (2) chaotic motion of the moduli in the
early universe.  To fix ideas we examine in detail the example of the
(conjectural) dilaton-axion ``$S$-duality'' of four-dimensional string
compactifications. The facts (1) and (2) together might help to solve
some problems with the standard scenarios for supersymmetry breaking
and vacuum selection in string theory.

\Date{March 9, 1994}
%\draft

\newsec{Duality in string theory}

Many moduli spaces of vacua in string theory are of the form
\eqn\emodspcs{
\CM = K ~ \backslash ~ \CG ~ / ~ \Gamma
}
where $\CG$ is a noncompact semisimple Lie group, $K\subset \CG$ is a
compact subgroup, and $\Gamma\subset \CG$ is a discrete subgroup. From a
physical point of view $K\backslash \CG$ is a moduli space of nonlinear
sigma model actions. When the spacetime fields of the action are
related by the discrete group $\Gamma$ the sigma models describe
identical conformal field theories. This phenomenon is referred to as
``$T$-duality.'' (See \giveon\ for a recent review.)  These spaces are
noncompact, the ``ends'' corresponding to weak coupling or
decompactification limits. Nevertheless, for some purposes, the spaces
$\CM$ behave like compact spaces. For example, in all orbifold and
toroidal compactifications the subgroup $\Gamma\subset \CG$ turns out to
be a {\it lattice} subgroup.  This means that the quotient $\CG/\Gamma$
has {\it finite volume} in the Haar measure. Furthermore, the typical
geodesic flow on spaces of the form $\CM$ is {\it
chaotic}~\refs{\mautner,\ccmr}.

The above facts have played a role in investigations of fundamental
symmetry properties of strings~\finite.  The purpose of this note is
to remark that these general facts also have profound implications for
the behavior of string moduli in the early universe. We will
illustrate this remark by focusing on one special case, namely the
behavior of the axion-dilaton modulus (the
``axiodil''\foot{{\bf ,}ak-s\={e}-{\bf '}yo-dil}) in four-dimensional
string compactifications.  Recall that in four dimensions (and only in
four dimensions), the naturally occurring three-form ${\bf H}$ can be
replaced by a scalar field $a$ (the ``model independent axion'') using
\eqn\eaxion{
H_{\mu\nu\lambda} = {1 \over 2 \kappa} e^{4 \phi}
 \epsilon_{\mu\nu\lambda\rho} \partial^\rho a \; ,
}
where $\phi$ is the string dilaton and $\kappa^2 = 8 \pi G$. The axion
can then be combined with $\phi$ into a single complex scalar field
\eqn\etau{
\tau \equiv a + i e^{-2 \phi} =
{ \theta \over 2 \pi } + \alpha^{-1} i \equiv x + i \, y\; ,
}
where $\theta$ is the vacuum angle, $\alpha$ is the string coupling
constant, and $x,y \in \IR$. With this definition, the low energy
effective action of string theory (in the Einstein metric) is
\eqn\eaction{\eqalign{
I = {1 \over 2 \kappa^2} \int \! {\rm d}^4 x \sqrt{-g} \bigg[ &
R
- {1 \over 2} { \partial_\mu \tau \partial^\mu \bar\tau \over (\Im \tau)^2}
- \kappa^2 V(\tau) \cr
& - {\kappa^2 \over 4 \pi} \Tr \left(
a F_{\mu\nu} \tilde{F}^{\mu\nu}
 - e^{-2 \phi} F_{\mu\nu} F^{\mu\nu} \right) +\cdots \bigg] \; , \cr
}}
where $R$ is the scalar curvature, $V(\tau)$ is a potential for
$\tau$, $F_{\mu\nu}$ is the Yang-Mills field strength, $\tilde{F}_{\mu\nu}
= {1 \over 2} \epsilon_{\mu\nu\lambda\rho} F^{\lambda\rho}$, and $\Tr
(t_a t_b) = - \half \delta_{ab}$. Note that ${\Im \tau >0}$, that is,
$\tau$ takes its values in the upper half plane $H=SO(2)\backslash
SL(2,\IR)$.  Moreover, the kinetic term for the axiodil defines a
sigma-model on $H$ with the Poincar\'e metric.

The duality symmetry we will explore is known as
``$S$-duality''\nref\font{A.~Font, L.E.~Ib\'{a}\~{n}ez, D.~L\"{u}st,
and F.~Quevedo, ``Strong-Weak Coupling Duality and Nonperturbative
Effects in String Theory,'' {\it Phys. Lett.} {\bf B249} (1990)
35.}\nref\rey{S.J.~Rey, ``The Confining Phase of Superstrings and
Axionic Strings,'' {\it Phys. Rev.} {\bf D43} (1991)
526.}\nref\schwarza{J.H.~Schwarz and A.~Sen, ``Duality Symmetric
Actions,'' {\it Nucl. Phys.} {\bf B411} (1994) 35,
hep-th/9304154.}\nref\schwarzb{J.H.~Schwarz and A.~Sen, ``Duality
Symmetries of 4-D Heterotic Strings,'' {\it Phys. Lett.} {\bf B312}
(1993) 105, hep-th/9305185.}\nref\schwarzc{J.H.~Schwarz, ``Does String
Theory have a Duality Symmetry Relating Weak and Strong Couplings?''
July 1993, hep-th/9307121.}\nref\sen{A.~Sen, ``Strong-Weak Coupling
Duality in Four Dimensional String Theory,'' Feb.~1994,
hep-th/9402002.}~\refs{\font {--} \sen}. The $S$-duality group is
$\Gamma=PSL(2,\IZ)$ and acts on $\tau$ via fractional linear
transformations:
\eqn\eslz{
\tau \to { a \tau + b \over c \tau + d} \;\;
{\rm where} \;\; a,b,c,d \in \IZ, a d - b c = 1 \>.
}
If the gauge group is abelian a suitable transformation law for
$A_\mu^I$ renders the entire action~\eaction\ $S$-duality
invariant~\refs{\schwarza{--}\schwarzc}.  If $\Gamma$ is a symmetry of
the theory then we learn that the axiodil is in fact described by a
nonlinear sigma model with target space
\eqn\modcurv{
\CF=SO(2)\backslash SL(2,\IR)/SL(2,\IZ)
}
which may be thought of as the famous keyhole region in the upper
half-plane with boundaries suitably identified.  It is important to
stress that while $T$-duality is a well established fact of string
theory, $S$-duality remains conjectural.\foot{The strongest evidence
for $S$-duality comes from backgrounds with $N=4$ supersymmetry.}
Nevertheless, we focus on this example because it is conceptually
simple, and because the consequences of such a symmetry would be
dramatic.

In section~2 we describe in more detail how the restriction of $\tau$
to $\CF$ implies that the simplest time development of $\tau$ is
chaotic.  In order to explore the implications of this fact for the
early universe we must form some idea of what nonperturbative
potential might be generated for the axiodil.  In section~3, we will
show how modular invariance greatly constrains the form of $V(\tau)$.
A few simple examples of allowed superpotentials are presented in
section~4.  The space \modcurv\ has a finite volume, $\pi/3$.  In
section five we argue that this implies that in the presence of
chaotic motion some problems with existing models of supersymmetry
breaking are naturally eliminated.  We conclude in section 6.

\newsec{The ergodic axiodil}

In the very early universe there is no potential for the axiodil
(above the scale of gaugino condensation) or the kinetic energy of
$\tau$ is so large that the potential energy can be ignored.  Under
these circumstances the time-development of the axiodil is very
simple.  Let us assume that $\tau(x^\mu)$ is spatially homogeneous in
an expanding Friedmann-Robertson-Walker spacetime with scale factor
$R(t)$. Hence $\tau$ is described by a single mode $\tau(t)$.
Moreover, as long as the volume of the universe is large in Planck
units (so that $\hbar/V \ll \hbar$), we describe the development of
$\tau(t)$ as the motion of a {\it classical} particle.

Neglecting the other fields and the potential $V$, the equation of
motion for $\tau$ and Einstein's equation derived from the
action~\eaction\ are
\eqn\eeomgen{\eqalign{
\nabla^2 \tau + { i (\nabla \tau)^2 \over \Im \tau} & = 0 \; ,\cr
R_{ab} - {1 \over 2} {\p_a \tau \p_b \bar{\tau} \over (\Im \tau)^2} & = 0 \;
.\cr
}}
Specializing to a flat FRW universe with spatially homogeneous $\tau$
reduces \eeomgen\ to
\eqna\mttau
$$\eqalignno{
\ddot\tau + {i\over \Im\tau} \dot\tau^2 + 3 H \dot \tau&=0 \; ,&\mttau{a}\cr
H^2 = {1 \over 12} {|\dot \tau|^2\over (\Im\tau)^2} \; ,&&\mttau{b}\cr
\dot H + 3 H^2 = 0 \; , &&\mttau{c}\cr
}$$
where $H=\dot R/R$. If $M_{\rm pl}\to \infty$, we set $H=0$ in
\mttau{a}.

It is easy to find the most general solution of \mttau{}. The
equations are $SL(2,\IR)$ invariant. Moreover, if one makes the ansatz
$\tau(t)=i e^{f(t)}$, then \mttau{} reduces to ${\ddot f + 3 H \dot f=0}
\Rightarrow \dot f(t) \propto1/R^3(t) \Rightarrow f(t)=f(0)\pm {2\over
\sqrt{3}} \log\left(1 + {t\over t_0}\right)$. Applying an $SL(2,\IR)$
transformation to this solution we obtain the most general solution:
\eqna\gensol
$$\eqalignno{
\tau(t) & = a + r e^{i \phi(t)} &\gensol{a}\cr
e^{i \phi(t)}&= {i \beta(t) -1 \over i \beta(t) + 1} &\gensol{b} \cr
\beta(t)&=\beta(0) \biggl(1+t/t_0 \biggr)^{\pm{2\over \sqrt{3}}}
   &\gensol{c}\cr
}$$
where $0\leq t < +\infty$, $a\in\IR$, $0<t_0, r,\beta(0)<\infty$.
Eq.~\gensol{} is the most general solution since there are four
independent parameters. The initial kinetic energy determines
${E(0)=1/(9 t_0^2)}$, the slope of the initial velocity determines
$\cot(\phi(0))=-\dot y/\dot x$, and hence fixes $\beta(0)=\cot(\half
\phi(0))$. The initial position then fixes $r$ and $a$. In the limit
$M_{\rm pl}\to\infty$, \gensol{} still holds except \gensol{c} becomes
\eqn\nograv{
\beta(t) = \beta(0) e^{\pm{2\over \sqrt{3}} (t/t_0)} \; .
}

It follows from \gensol{} and \nograv\ that in the absence of gravity
the trajectory of $\tau$ in the upper half plane is a semicircle of
radius $r$ centered at $a$. In the presence of gravity the Hubble
friction affects the rate at which the semicircle is traversed and
restricts the trajectory to an arc with one endpoint on the $x$-axis.

It is well-known that motion on compact manifolds of constant negative
curvature is chaotic. Indeed, in the hierarchy of chaotic systems it
is the ``most chaotic'' type of dynamical system. More technically,
these dynamical systems are ``K-systems.'' Roughly speaking, this
means that there exists a timescale, $t_0$, such that a typical motion
samples the entire space on the timescale $t_0$ and such that nearby
initial conditions lead to trajectories which diverge on timescale
$t_0$ (i.e., $1/t_0$ is a Lyapunov exponent).
See\nref\arnold{V.I.~Arnold, {\it Ergodic Problems of Classical
Mechanics} (WA Benjamin, New York, 1968).}\nref\sinai{I.P.~Cornfeld,
S.V.~Fomin, and Y.G.~Sinai, {\it Ergodic Theory} (Springer-Verlag, New
York, 1982).}\nref\anosov{D.V.~Anosov, ``Geodesic Flows on Closed
Riemann Manifolds with Negative Curvature,'' in {\it Proc.~Steklov
Inst.~Math.}~90.}\nref\zaslovsky{G.M.~Zaslovsky, {\it Chaos in Dynamic
Systems} (Harwood, New York, 1985).}\nref\gutzwiller{M.C.~Gutzwiller,
{\it Chaos in Classical and Quantum Mechanics} (Springer-Verlag, New
York, 1990).}~\refs{\arnold {--} \gutzwiller} for details. In the
present case although $\CF=SO(2)\backslash SL(2,\IR)/SL(2,\IZ)$ is
noncompact (because of the region at weak coupling), in many ways it
behaves like a compact space. For example, $\CF$ has finite volume in
the Poincar\'e metric. The motion described by \nograv\ is still
chaotic. Indeed, it is still a K-system~\sinai.\foot{For an
explanation of the ergodicity, which goes back to J.~Hadamard and
E.~Artin, in the string literature see, e.g., \finite.}

\ifig\funnel{Part of a typical geodesic of the axiodil on $\CF$.}
{\epsfxsize=320pt\epsfbox{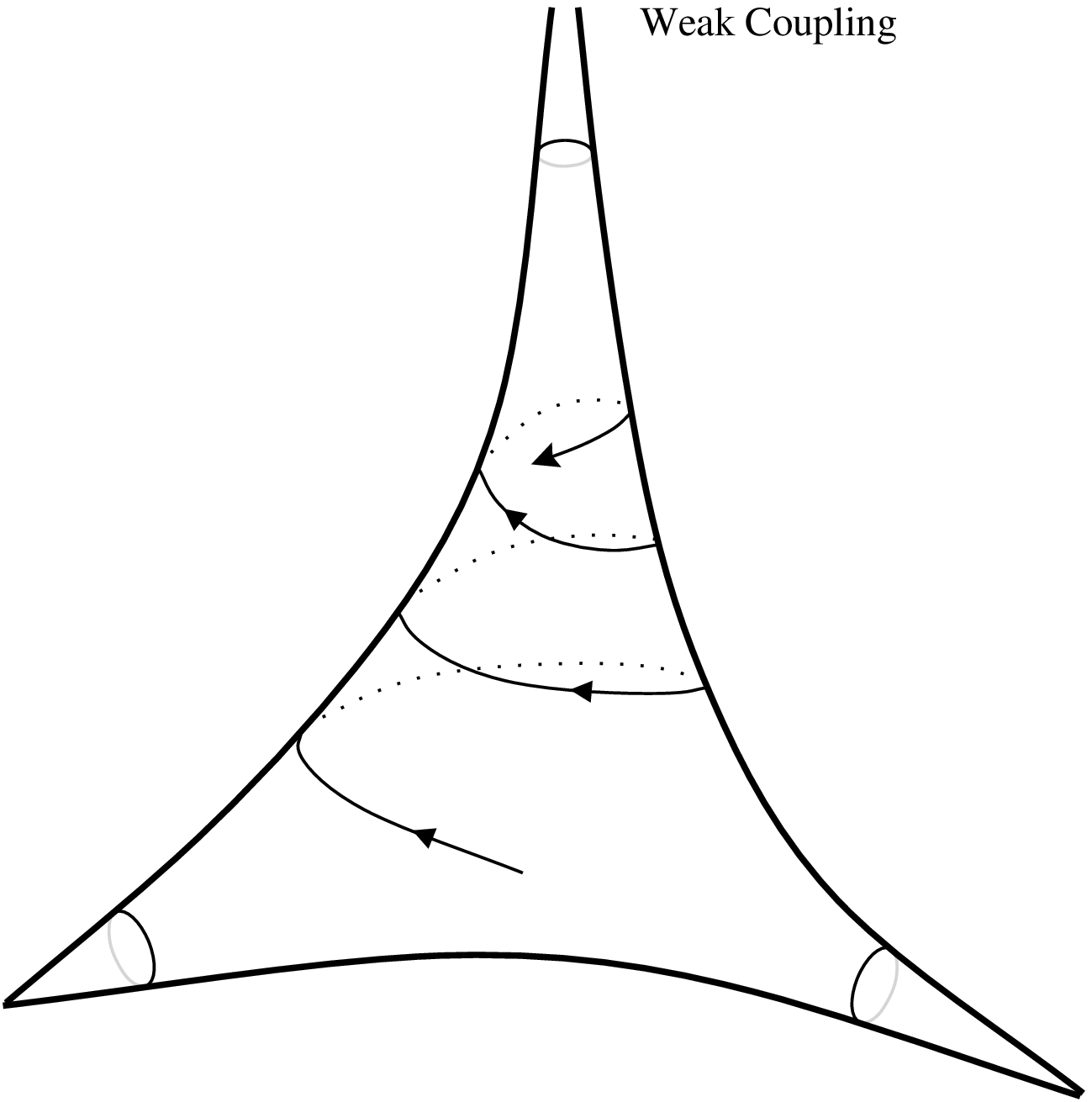}}

It is useful to understand the general nature of the motion of
$\tau(t)$ in heuristic terms.  As the semicircle approaches the
$x$-axis the trajectory passes through an infinite number of
fundamental domains.  Mapping the semicircle trajectory back into the
fundamental domain one sees that $\tau$ spirals out to weak coupling,
bounces back (when it reaches the crest of the semicircle), then
bounces in a ``random'' way around the region of strong coupling, then
spirals out to weak coupling, then bounces back again, and so on. The
result is, typically, a dense orbit. We have illustrated this in
\funnel.

In the presence of gravity the trajectory~\gensol{} of the axiodil in
the upper half plane still lies along a semicircle. Since one end of
the arc descends to the $x$-axis we can use the standard result about
chaotic motion of geodesics on $\CF$ to conclude that the motion of
$\tau$ in the early universe is chaotic, albeit on logarithmically
changing timescales.

\newsec{$S$-Dual Potentials}

In order to study motion when $V\not=0$ we first discuss some
general aspects of the kinds of potentials we expect.

\subsec{Constraints on the superpotential $W(\tau)$}

As in the case of $T$-duality\nref\fflst{S.~Ferrara, D.~L\"{u}st,
A.~Shapere, and S.~Theisen, ``Modular Invariance in Supersymmetric
Field Theories,'' {\it Phys. Lett.} {\bf B225} (1989)
363.}\nref\filq{A.~Font, L.E.~Ib\'{a}\~{n}ez, D.~L\"{u}st and
F.~Quevedo, ``Supersymmetry Breaking from Duality Invariant Gaugino
Condensation,'' {\it Phys. Lett.} {\bf B245} (1990) 401;
M.~Cveti\v{c}, A.~Font, L.E.~Ib\'{a}\~{n}ez, D.~L\"{u}st and
F.~Quevedo, ``Target Space Duality, Supersymmetry Breaking and the
Stability of Classical String Vacua,'' {\it Nucl. Phys.} {\bf B361}
(1991) 194.}~\refs{\fflst,\filq}, $S$-duality strongly restricts the
form of the potentials that can appear in the effective action~\font.
Since the effective potential for moduli in the effective low-energy
$N=1$, $D=4$ supergravity theory arises entirely from uncalculable
nonperturbative effects this is useful information. In this subsection
we isolate the general class of potentials allowed by

\noindent
{\bf Condition~1:} Validity of weak coupling perturbation theory.

and

\noindent
{\bf Condition~2:} $S$-duality.

We impose condition~1 since, after all, the only truly solid basis for
string theory is its perturbative series! Weak coupling corresponds to
$\Im \tau \to \infty$ or $q \equiv e^{2 \pi i \tau}\to 0$. Our
discussion will be subject to several caveats listed at the end of
this section.

We ignore other moduli and concentrate on the dynamics of the
axiodil alone. The potential of $N=1$ supergravity has the
form~\cjsf
\eqn\esuperv{
V(\tau) = e^{G(\tau)} \left[
{ | \partial_\tau G(\tau) |^2
\over \partial_\tau \partial_{\bar\tau} G(\tau) } - 3 \right] \; ,
}
where $G(\tau)$ is given by
\eqn\esuperg{\eqalign{
G(\tau) &=K(\tau,\bar{\tau}) + \log | W(\tau) |^2 \cr
&= - \log(\Im \tau) + \log | W(\tau) |^2 \cr}
}
and hence:
\eqn\esuperw{
V(\tau) = { |W(\tau)|^2 \over \Im \tau }
\left( | 1 - 2 i \Im \tau \partial_\tau \log W |^2 - 3 \right) \; .
}

We first impose $S$-duality. This symmetry implies that $\tau$ is a
coordinate on the space $\CF=SO(2)\backslash SL(2,\IR)/SL(2,\IZ)$. It
is a general result that the superpotential $W$ must be a section of a
holomorphic line bundle $\CL\to\CF$~\ref\bggrwit{J.~Bagger and
E.~Witten, ``Quantization of Newton's Constant in Certain Supergravity
Theories,'' {\it Phys. Lett.} {\bf B115} (1982) 202.}. The Chern class
of $\CL$ is determined by the metric on $\CL$ which follows from $K$:
$\parallel W \parallel^2=e^{G}$. It follows that $W$ can be written
as:
\eqn\ewform{
W(\tau) = { f(J(\tau)) \over \eta(\tau)^2 } \; ,
}
where $\eta(\tau)$ is the Dedekind $\eta$-function, $J(\tau)$ is the
generator of modular functions of weight~0,\foot{Some relevant facts
about $J$ are collected in appendix~A.} and $f$ defines a section of a
flat holomorphic line bundle over $\CF$.

The group of flat holomorphic line bundles over $\CF$ is
$\IZ/6\IZ$\nref\mumf{D. Mumford, ``Picard Groups of Moduli Problems,''
in {\it Arithmetical Algebraic Geometry} (O.F.G. Schilling, ed.,
Harper and Row, New York, 1965).}\nref\lehner{J. Lehner, {\it
Discontinuous Groups and Automorphic Functions} (Amer. Math. Soc.,
Providence, 1964).}~\refs{\mumf,\lehner}. A section of a flat
holomorphic bundle satisfies the modular transformation law $f(J) \to
e^{i \phi}f(J) $ where $e^{i \phi}$ is a one-dimensional
representation of the modular group. The modular group may be
presented in terms of generators and relations as: $PSL(2,\IZ)=\langle
S,T|S^2=1, (ST)^3=1\rangle $ so $S,T$ are represented by complex
numbers $\CS,\CT$ which must satisfy: $\CS^2=\CS^3 \CT^3=1$. It
follows that $\CS$ is a square root of unity and $\CT$ is a sixth root
of unity. The orbifold points of $ST,S$ are $\tau=e^{ 2\pi i/3},i$
of order $3,2$, respectively, so that if $f$ transforms nontrivially
under $ST,S$ it must either vanish or have a pole at $\tau=e^{2\pi
i/3},i$. Thus the general form of $f$ following from $S$-duality is
$f(J)=J^{n/3}(J-1728)^{m/2} R(J)$ where $R(J)$ is a rational function
of $J$ and $n,m$ are integral. Note that the behavior of $f$ under
$\tau\to\tau+1$ near $\tau \sim i \infty$ fixes $n \mod 3, m \mod 2$.

Now we insist that weak coupling perturbation theory is a good
approximation. In particular $W\to 0$ for $q\to 0$. Since $J\sim 1/q$
for $q\to 0$, we deduce that the general form of the superpotential
must be
\eqn\ewformi{
W(\tau) = { 1 \over
\eta(\tau)^2 } J^{n/3}(J-1728)^{m/2} {P_1(J)\over P_2(J)} \; ,
}
where $n,m \in \IZ$, the $P_i$ are polynomials, and
$\deg P_2> \deg P_1 + n/3 + m/2 + 1/12$.
Duality invariant potentials have been discussed in
\refs{\fflst,\filq} for $T$-duality and in \font\ for $S$-duality, but
without imposing condition~1. Instead, \font\ imposed the condition
that the superpotentials be regular everywhere except for the $q \to
0$ limit. This condition forces the superpotential to be singular in
the weak coupling limit, and not vanishing. Thus the allowed
potentials in \font\ are different from the general form~\ewformi.

\subsec{Application: Gaugino Condensation}

It is widely believed that the source of the effective potential for
$\tau$ (and other moduli) which will stabilize the dilaton, and,
perhaps, break supersymmetry lies in gaugino condensation in a
strongly interacting hidden gauge sector\nref\derend{J.P.~Derendinger,
L.E.~Ib\'{a}\~{n}ez, and H.P.~Nilles, ``On the Low-Energy $D=4$, $N=1$
Supergravity Theory Extracted from the $D=10$, $N=1$ Superstring,''
{\it Phys. Lett.} {\bf B155} (1985) 65.}\nref\drsw{M.~Dine, R.~Rohm,
N.~Seiberg, and E.~Witten, ``Gluino Condensation in Superstring
Models,'' {\it Phys. Lett.} {\bf B156} (1985)
55.}\nref\krasnikov{N.V.~Krasnikov, ``On Supersymmetry Breaking in
Superstring Theories,'' {\it Phys. Lett.} {\bf B193} (1987)
37.}\nref\dixon{L.J.~Dixon, ``Supersymmetry Breaking in String
Theory,'' SLAC-PUB-5229, in {\it Div.~of Particles and Fields
Conf.~Proc.}~(1990) 811.}\nref\nilles{H.P.~Nilles, ``Gaugino
Condensation and Supersymmetry Breakdown,'' {\it Int. Jour. Mod.
Phys.} {\bf A5} (1990) 4199.}\nref\louis{J. Louis, ``Non-Harmonic
Gauge Coupling Constants in Supersymmetry and Superstring Theory,''
{\it Proc. of the 1991 DPF meeting} (World Scientific, Singapore 1992)
751.}~\refs{\derend {--} \louis}. One interesting set of
models~\carlos\ is based on a hidden sector with gauge group $SU(N)_k$
where $k$ is the level of the affine Lie algebra that generates the
symmetry.  The hidden sector has matter consisting of $M$ quark
families and a gauge singlet $A$ coupled to the quarks via mass terms
and to itself via a tree-level superpotential $W\sim A^3$.  In
addition to the tree-level superpotential a nonperturbative
superpotential is generated by gaugino condensation.  Integrating out
$A$ leads to the effective superpotential (to leading order in $q$)
\eqn\ecasasone{
W(\tau) = d \, e^{ 6 \pi i \tau k/(3N - M)} = d \, q^{ 3k/(3N - M) },
}
where $d$ is independent of $\tau$, but depends on $M$, $N$, and the
compactification moduli.

Let us assume that $S$-duality is unbroken in the models of \carlos\
and that \ecasasone\ is the first term of a series which must be
promoted to an $S$-dual expression. Given the general form~\ewformi\
this can be done if and only if
\eqn\dioph{
{3k\over 3N-M} = -{1\over 12} + {\ell\over 6} \quad {\rm where} \; \ell>0 \; .
}
Writing $\ell=2n + 3m$, one natural
$S$-dual completion of \ecasasone\ is:
\eqn\promote{
W= {d\over \eta^2} J^{-n/3}(J-1728)^{-m/2} \; .
}
The higher powers of $q$ might arise from instanton sums or from
condensates of heavy fields or from both sources.

A potential based on \ecasasone\ cannot give a stable minimum to
$\tau$ at weak coupling\foot{Although an $S$-dual promotion can have a
minimum at strong coupling. See the next section.} so models with two
or more condensates (``racetrack models'') are usually
considered~\refs{\krasnikov,\dixon,\carlos}. They have superpotentials
of the form
\eqn\ecasastwo{
W(q) = d_1 q^{ 3 k_1/(3 N_1 - M_1)} + d_2 q^{ 3 k_2/(3 N_2 - M_2) } \; .
}
A potential of this type can also be easily accommodated in a manner
that preserves $S$-duality as in \promote:
\eqn\efexmpa{
f(J) = d_1 J^{ - n_1/3} (J-1728)^{-m_1/2} + d_2 J^{ - n_2/3}
(J-1728)^{-m_2/2} \; .
}
Note that this can only agree with the general expression~\ewformi\ if
in addition to \dioph\ we have
\eqn\etworest{
{3 k_1 \over 3 N_1 - M_1} - {3 k_2 \over 3 N_2 - M_2} \in \IZ \; .
}
In particular, a minimum at weak coupling
only exists when the ratio $d_1/d_2$ is very large or small.

In a four-dimensional heterotic string compactification, the total central
charge of the hidden sector must satisfy
\eqn\ecenc{
c_{\rm hidden} = \sum_{G_i} c_i = \sum_i { k_i (N_i^2 - 1) \over k_i + N_i}
 \le 18 \; ,
}
where the sum is over the different gauge groups. The
constraint~\ecenc\ for a single condensate can be combined with the
$S$-duality constraint~\dioph\ to limit the possible $S$-dual gaugino
condensates. The various possible one condensate models can then be
combined in a variety of ways to form two condensate models, subject
to the constraints~\dioph, \etworest, and \ecenc. Note that
$S$-duality is incompatible with all the two condensate models of
\carlos. Evidently, unbroken $S$-duality is a powerful principle!

\noindent
{\bf Warnings and Caveats}

\noindent
1. We have assumed that it is reasonable to consider the axiodil
$\tau$ in the absence of other moduli. It is not clear that this is
reasonable. As has been stressed in \sen\ it is impossible, with our
current understanding, to prove or disprove $S$-duality by examining
quantities which are corrected in string perturbation theory. It might
well be that one must transform all the moduli together \sen.  Indeed,
the 1-loop correction to the K\"ahler potential for
$\tau$\nref\dfkz{J.P.~Derendinger, S.~Ferrara, C.~Kounnas, and
F.~Zwirner, ``On Loop Corrections to String Effective Field Theories:
Field Dependent Gauge Couplings and Sigma Model Anomalies,'' {\it
Nucl. Phys.} {\bf B372} (1992) 145.}\nref\anton{I.~Antoniadis,
E.~Gava, K.S.~Narain, and T.R.~Taylor, ``Superstring Threshold
Corrections to Yukawa Couplings,'' {\it Nucl. Phys.} {\bf B407} (1993)
706, hep-th/9212045.}~\refs{\dfkz,\anton}\ mixes $\tau$ with other
moduli in a nontrivial way, is not obviously $S$-dual and, unlike the
superpotential, does not have a natural $S$-dual nonperturbative
extension.  These results do not disprove, but do weaken the case for
$S$-duality in the absence of $N=4$ spacetime supersymmetry.

\noindent
2. The constraint \etworest\ can be violated by the existence of vev's
of heavy string modes which transform under the $S$-duality
group.\foot{This possibility was pointed out to us by T.~Banks and
N.~Seiberg.} For example, it might be that a relative phase between
the two terms in \efexmpa\ under $\tau\to \tau +1$ merely indicates
that some massive string mode gets a vev spontaneously breaking
$S$-duality: $d_2\sim \langle \omega\rangle$ where $\omega$ is a heavy
field transforming as $\omega\to \omega e^{2\pi i
(\alpha_2-\alpha_1)}$ under $\tau\to\tau+1$. An example of such
phenomena can be found in \ref\irrtax{T.~Banks, M.~Dine, and
N.~Seiberg, ``Irrational Axions as a solution of the strong CP problem
in an eternal universe,'' {\it Phys. Lett.} {\bf B273} (1991) 105,
hep-th/9109040.}.

\newsec{Some examples of the potential}

We will examine two simple cases of potentials of the type isolated in
the previous section.

\subsec{$J^{-1/3}$ potential}

One of the simplest superpotentials has $f(J) = 1/J^{1/3}$, or
\eqn\ewone{
W(\tau) = {1 \over \eta(\tau)^2 J(\tau)^{1/3} } \; .
}
With a bit of algebra, we can derive a useful formula for $V(\tau)$.
First, define
\eqn\eetilde{
\tilde{E}_4(q) = {3 \over 2 \pi i} \partial_\tau E_4(q) =
 720 \sum_{n = 1}^{\infty} n \sigma_3(n) q^n \; ,
}
and split $\eta(\tau)$ into its fractional power and its integer
powers using
\eqn\eetab{
\eta(q) = q^{1/24} \prod_{n = 1}^{\infty}{ (1 - q^n)}
\equiv q^{1/24} \tilde{\eta}(q) \; .
}
Now some algebra reduces $V(\tau)$ from \esuperw\ to
\eqn\evone{
V(\tau) = { e^{ - \pi y} |\tilde{\eta}(q)|^{12}
 \over y |E_4(q)|^4 } \left[ \left| E_4(q) +
 {\pi \over 3} y \left( 3 E_2(q) E_4(q) -
 4 \tilde{E}_4(q) \right) \right|^2 - 3 |E_4(q)|^2 \right] \;
}
(remember that $y = \Im \tau$).

The motivation for this form of $V(\tau)$ is as follows: the upper
half plane contains an infinite number of copies of the fundamental
domain $\CF$. We are free to choose the standard fundamental region $D
= \{ z| \Im z > 0, \Re z \le \half, |z| \ge 1\}$ (the ``keyhole''
domain), which has corners at $\rho \equiv e^{i \pi/3} \sim e^{2 i
\pi/3}$. In the standard fundamental region $D$, $q = e^{2 \pi i
\tau}$ is a small parameter, and $q$ shrinks rapidly as $\Im \tau \to
\infty$. The functions $\tilde{\eta}(q)$, $E_2(q)$, $E_4(q)$, and
$\tilde{E}_4(q)$ all have Taylor series expansions around $q = 0$ that
converge for all of $D$, as can easily be seen from the equations in
the appendix and eqs.~\eetilde\ and~\eetab. The radius of convergence
in all cases is $|q| = 1$, or $y = 0$, which is well outside
$D$~\serre. Thus, $V(\tau)$, as written in \evone, can be expressed as
the ratio of two Taylor expansions, each carried out to some desired
degree of accuracy by including sufficiently high order powers of $q$.
None of the functions $\tilde{\eta}(q)$, $E_2(q)$, $E_4(q)$, or
$\tilde{E}_4(q)$ have poles in $D$. The first order zero of $E_4(q)$
at $\tau = \rho$ causes $V(\tau)$ to have a pole at $\tau = \rho$,
which is why $V(\tau)$ must be calculated as the ratio of two
expansions. Since $E_4(q)$ is the only function in the denominator of
$V(\tau)$, this is the only pole of $V(\tau)$.

\ifig\fvone{A contour plot of $\log (V(\tau) - V_{\rm min})$ with $f(J)
= 1/J^{1/3}$. Darker shades mean lower values. There is a pole at $\tau =
\rho$ and a global minimum $V_{\rm min}$ at $\tau = i$.
$V(\tau) \to 0$ as $y \rightarrow \infty$. The thick line at
$|z| = 1$ marks the boundary of the fundamental region~$D$.}
{\epsfbox{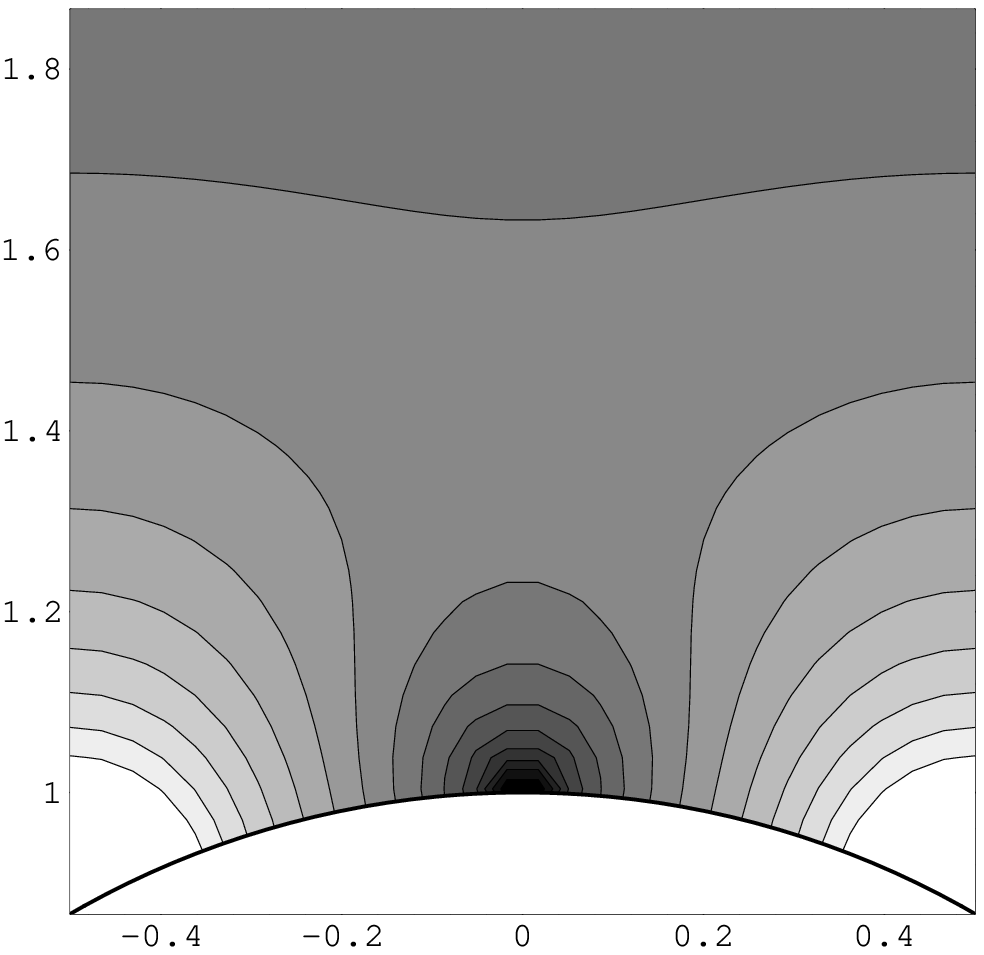}}

A plot of $V(\tau)$ is shown in \fvone. As mentioned, it has a pole at
$\tau = \rho$, the two corners of $D$. Because $e^{i \pi/3}$ is
identified with $e^{2 i \pi/3}$ in $D$, this is only one pole, not two
as it may appear.  The $\exp( -\pi y)$ factor causes $V(\tau)$ to fall
off rapidly away from the pole. For large $y$,
\eqn\evlarge{
V(\tau) \to \pi^2 y e^{- \pi y} \; ,
}
so the potential does vanish at weak coupling, as desired. $V(\tau)$
also has a minimum at $\tau = i$. This is actually the global minimum,
since the first absolute value in \evone\ vanishes there, causing
$V(\tau)$ to become negative in a small region near $\tau = i$. To be
precise, $V(\tau = i) = -5.98142 \times 10^{-2}$. This region is
separated by a potential barrier from the region towards large $\Im
\tau$. These are the main features of the potential.

The numerics of the calculation of $V(\tau)$ in \fvone\ are quite
good. The potential in \fvone\ was produced by keeping the first~20
terms in $q$ in the power series expansion of both the numerator and
denominator of \evone. The main source of numerical error is in the
precise position of the pole. We choose to include the first 20
terms so that the position of the pole is correct to $10^{-5}$.
Stability improves instantly away from the pole. Including~21 terms
in the $q$ expansions changes $V(\tau)$ by at most one part in
$10^{16}$ for $y < 1$ and by even less as $y$ increases. Recall that
$q^{20} = 2.66 \times 10^{-55}$ for $y = 1$. The extra accuracy gained
by including the twenty-first term at $\tau = i$ is one part in
$10^{23}$.

Let us examine the physical implications of \evone. Consider an
expanding and cooling universe which develops a potential~\evone. As
the axiodil cools, its energy drops below the level of the barrier
separating the minimum at $\tau = i$ and the region near $\infty$
(more about this in the next section). If it falls outside the region
near $\tau = i$, it will continue rolling out to $\infty$. Otherwise,
$\tau$ will eventually settle into its minimum at $\tau = i$, or
$\theta = 0$ and $\alpha^{-1} = e^{-2 \phi} = 1$. This minimum has two
properties that are in conflict with the observed universe. First, the
effective axiodil vacuum energy at $\tau=i$ is $-5.98142 \times
10^{-2} M_{\rm pl}^{4}$. Since other moduli also contribute effective
vacuum energies, this is not necessarily a fatal flaw. Second, recent
experimental results indicate that $\alpha^{-1} \approx 25$ at the GUT
scale (see, for example, ref.~\langacker), and so physics should be
weakly coupled, not strongly coupled ($\alpha^{-1} = 1$) as this
potential predicts.  Finally, at this minimum $\p_\tau G=0$ so
supersymmetry must be broken by other fields, e.g., the $F$-terms of
compactification moduli.

\subsec{A ``two condensate'' model}

As discussed in the previous section, a model with two condensates
will have a superpotential of the form
\eqn\ewtwo{
W(\tau) = { c \over \eta(\tau)^2 J(\tau)^{n/3} (J(\tau)-1728)^{m/2}}
\left(1 + {\beta \over J(\tau)^p} \right)\; ,
}
where ${n \over 3} + {m \over 2} - {1 \over 12} > 0$, $p \in \IZ_+$,
and $c,\beta$ are (possibly complex) constants, and for the purpose at
hand we can take $c=1$. To simplify notation, define $r \equiv 4 n + 6
m - 1$ and $\tilde{E}_6(q) = {1 \over 2 \pi i} \p_\tau E_6(q)$. Some
work then reveals that
\eqn\evtwo{\eqalign{
V(\tau) = & { e^{- {r \over 3} \pi y} |\tilde{\eta}(q)|^{4 r}
\over y |E_4(q)|^{2 (n + 3p + 1)} |E_6(q)|^{2 (m + 1)}} \times \cr
 & \bigg\{ \bigg|
E_4^{3 p} \left[ E_4 E_6 + {\textstyle{\pi \over 3}} y \left( r
 E_2 E_4 E_6
- 4 n \tilde{E}_4 E_6 - 12 m E_4 \tilde{E}_6 \right) \right] + \cr
\beta q^p & \tilde{\eta}^{24 p} \left[ E_4 E_6 +
{\textstyle{\pi \over 3}} y \left( (r + 12 p ) E_2 E_4 E_6 -
4 (n + 3 p) \tilde{E}_4 E_6 - 12 m E_4 \tilde{E}_6 \right)\right]\bigg|^2 \cr
 & - 3 \left| E_4 E_6 \left( E_4^{3 p} +
\beta q^p \tilde{\eta}^{24 p}\right)\right|^2 \bigg\} \; . \cr
}}

Although \evtwo\ seems complicated, in practice its salient features
can be determined with only a little more work. For simplicity, for
the rest of this section set $p = 1$, and take $\beta$ to be real and
positive. To agree with experiment, the potential should have a
minimum with $\alpha^{-1} = y > 5$ or so. Thus, ${q = e^{2 \pi i x -
2\pi y}}$ will be extremely small near the minimum and most of the
higher order terms in \evtwo\ can be neglected. However, since $q$ is
small, to have a minimum $\beta$ must be approximately $1/q_{\rm
min}$, and this will be very large. A mechanism for producing large
values of $\beta$ was described in \carlos. We can expand \evtwo,
keeping only the lowest order terms in $q$, the next order terms in
$q$ if they contain $\beta$ (since we assume $\beta$ is large), and at
each level in $q$ keep only the terms with the highest power of $y$.
This gives
\eqn\evapprox{
V(\tau) \approx {r \over 3} \pi^2 y e^{ - {r \over 3} \pi y}
\left( {r \over 3} + 2 \left(4 + {r \over 3}\right)
\beta e^{-2 \pi y} \cos(2 \pi x) \right) \; .
}
This has a local minimum at $x = \pm {1 \over 2}$ and
\eqn\eyapprox{
y = {1 \over 2 \pi} \log \left( 2 \beta
\left(1 + {6 \over r}\right) \left(1 + {12 \over r}\right) \right) \; .
}
Thus, we see that for sufficiently large $\beta$, a local minimum
occurs at large $y$. Qualitatively, the basic picture of a two
condensate model is as follows: For large enough $\beta$, the second
term in \ewtwo\ will dominate for small $y$. Therefore, near the unit
circle the two condensate potential will behave just like a one
condensate potential. As $y$ increases, the second term in \ewtwo\
falls off more rapidly than the first, which eventually comes to
dominate when $\beta q^p \approx 1$. Since $\beta$ is positive, a
minimum occurs when $x = \pm {1 \over 2}$ which causes the two terms
to have opposite sign.

Let us now work out a specific example of a two gaugino condensate
model. In \ewtwo, take $c = 1$, $m = 0$, $n = 4$, $p = 1$, and $\beta
= 1.62 \times 10^{17}$. With these choices, the potential is
\eqn\evthree{\eqalign{
V(\tau) = & { e^{- 5 \pi y} |\tilde{\eta}(q)|^{60} \over y |E_4(q)|^{16}
}\bigg\{ \bigg|
E_4(q)^{3} \left[ E_4(q) + {\pi \over 3} y \left(15 E_2(q) E_4(q)
 - 16 \tilde{E}_4(q)\right) \right] \cr
 & + \beta q \tilde{\eta}(q)^{24} \left[ E_4(q) +
{\pi \over 3} y \left( 27 E_2(q) E_4(q) -
28 \tilde{E}_4(q)\right)\right]\bigg|^2 \cr
& - 3 \left| E_4(q) \left( E_4(q)^{3} +
\beta q \tilde{\eta}(q)^{24}\right)\right|^2 \bigg\} \; . \cr
}}
This choice of parameters corresponds to a two condensate model of the
type mentioned earlier in \ecasastwo, with one sector having $k_1 =
3$, $N_1 = 5$, $M_1 = 11$, and $d_1 = 7.56 \times 10^{14}$, and the
other sector having $k_2 = 5$, $N_2 = 4$, ${M_2 = 0}$, and ${d_2 = 4.66
\times 10^{-3}}$ (we have used the formula for $d_i$ from \carlos).
The approximation~\eyapprox\ indicates that this potential should have
a minimum at $y = 6.5$. This value of $\alpha_{\rm min}^{-1}$ is
somewhat lower than the $\alpha^{-1} \approx 25$ indicated by LEP, but
not enough to be completely ruled out.

\ifig\fvtwo{A contour plot of the potential~\evthree. The contour levels
are on a logarithmic scale, and darker regions mean lower values of
$V(\tau)$. The potential has a local minimum at $\tau_{\rm min} = \pm
{1 \over 2} + 6.3996 i$. The potential $V(\tau)$ is negative near
$\tau_{\rm min}$, but only for a region too small to be apparent.}
{\epsfbox{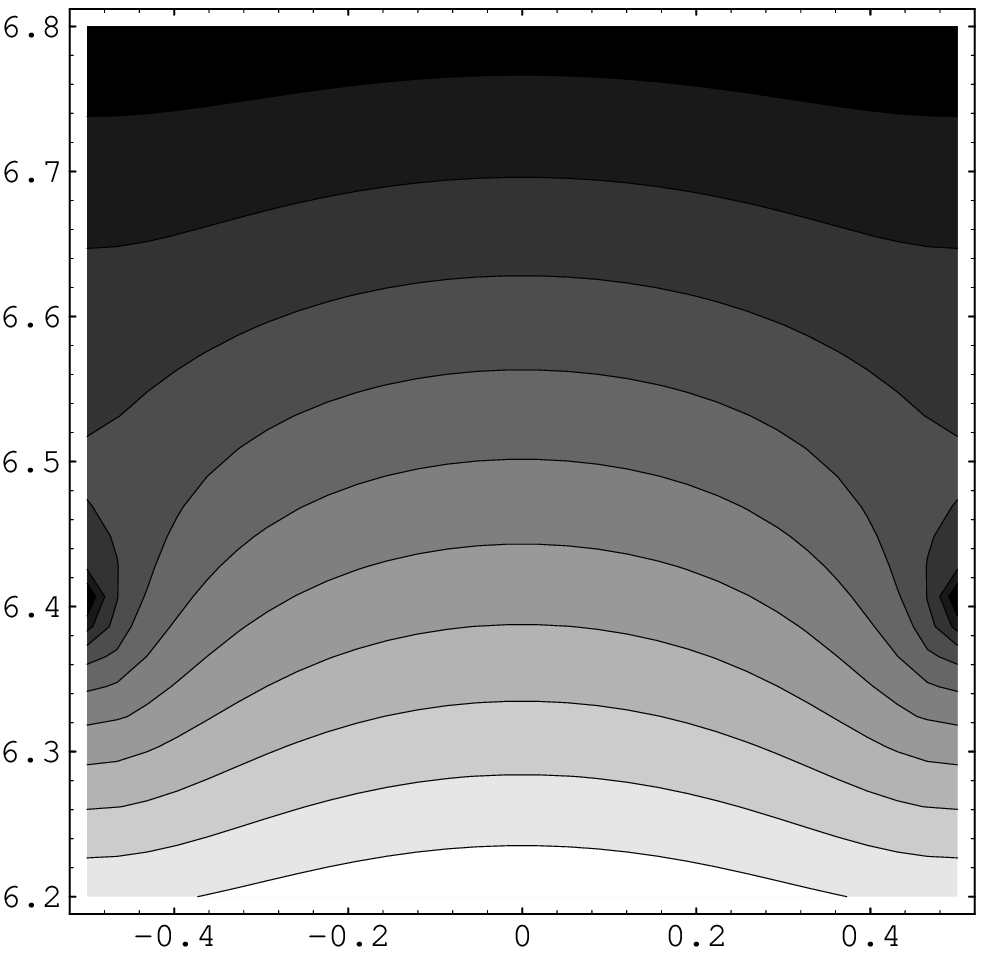}}

A contour plot (on a logarithmic scale) of the region of \evthree\
between $y = 6.2$ and $y = 6.8$ is shown in \fvtwo. The plot was
created by keeping the first ten terms in the $q$ expansion of
\evthree, but keeping even~20 terms in the $q$ expansion changes the
result by less than one part in $10^{18}$ at any point in the region
shown. The potential has a local minimum at $y = 6.3996$, $x = \pm {1
\over 2}$ (since, in $D$, $x = + \half$ is identified with $x = -
\half$, this is only a single minimum). Because $m = 0$, the region near
$y = 1$ (except for an overall scale) is similar to that for the
single condensate~\fvone, with a global minimum at $\tau = i$. Just
like at the minimum at $\tau = i$, the minimum of the
potential~\evthree\ at $y = 6.3996$ is negative, $V(\tau_{\rm min})
= -2.02 \times 10^{-45}$. Since we have not discussed the overall
scaling of $V(\tau)$, a more meaningful number is the ratio of the
value of $V(\tau)$ at this local minimum to that at the global minimum
at $\tau = i$. This ratio is
\eqn\evrat{
{V(\tau = \pm {1 \over 2} + 6.3996 i) \over V(\tau = i)}
= 1.15 \times 10^{-65} \; .
}
Thus it may be possible to say that even though $V(\tau_{\rm min})$ is
negative, it is not very negative.

\newsec{Problems solved by chaotic motion and $S$-duality}

\subsec{Fine Tuning}

One of the major objections to the standard string supersymmetry
breaking scenario is the problem of fine tuning of initial
conditions~\brustein. Since the potential has a variety of minima
(including the one at $\tau = + i \infty$), there is no {\it a priori}
reason for $\tau$ to fall into the desired minimum instead of falling
into one at strong coupling or rolling off towards $\infty$. Thus one
is forced to fine tune the initial conditions so that $\tau$ lies in
the desired minimum after supersymmetry breaking. The ideas of this
paper provide a potential solution to this problem.

\ifig\fyaxis{The potential~\evone\ along the $y$-axis.}
{\epsfbox{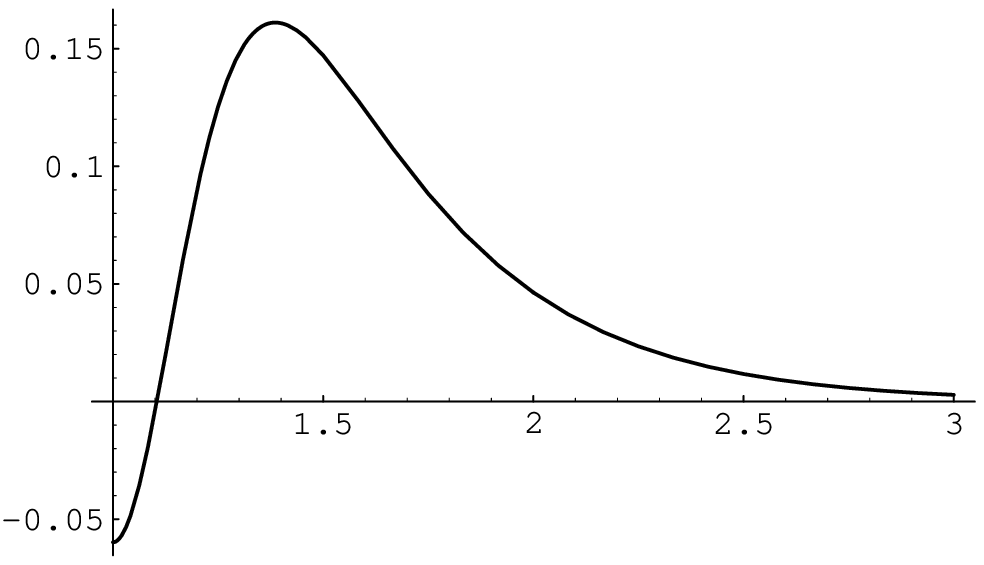}}

As discussed in section~2, in the very early universe $\tau$ moves
chaotically through the fundamental region. Thus, its initial
conditions will be erased after a few $e$-folding times, and the value
of $\tau$ at any time is best described by a probability distribution
inherited from the Poincar\'e metric. Thus, we may expect that the
probability of finding $\tau$ in a region $A$ is given by
\eqn\probdist{
P(A) = {3\over \pi} \int_A {{\rm d}x \, {\rm d}y\over y^2} \; .
}
This has the great virtue that {\it the region at weak coupling has
finite volume} so that the probability of getting trapped at a
nontrivial minimum can be nonvanishing.

Now imagine that a potential $V(\tau)$ forms and supersymmetry breaks.
To get a rough idea of what will happen we estimate that the
probability that $\tau$ will get trapped in a region containing a
certain minimum is given by the distribution~\probdist\ where we take
$A$ to be the basin of attraction for motion at small velocity for
that minimum. As an example the plot of $V(\tau)$ along the $y$-axis
is shown in \fyaxis\ for the case of a single condensate~\evone. If
$\tau$ is to the left of the maximum at $y \cong 1.4$, then it will
roll towards the minimum at $\tau = i$. Otherwise it will roll towards
$\infty$. Even though it may seem that just a small area of the
infinite region will fall into the minimum, the ${3\over \pi y^2}$
measure increases the probability that $\tau$ will be at low $y$, so
the probability that $\tau$ will fall into the minimum near $\tau = i$
is approximately 14\%. Thus the probability, though not unity, is
certainly not a small number,
% $>$ the chance of having a grant approved,
% $\gg$ the chance of getting a job
and no fine tuning is necessary.

The two condensate example~\evthree\ is more complicated, but an
optimistic estimate for the probability of settling into the weak
coupling minimum can be made as follows. First, the basin around
minimum of \evthree\ at $\tau = i$ is smaller than that of \evone, so
there is a 6\% chance that $\tau$ will fall into that minimum. Second,
14.7\% of the time $\tau$ will have $y > 6.5$, and so will not fall
into the minimum at $\tau_{\rm min} = \pm {1 \over 2} + 6.3996 i$.
About .03\% of the time, $\tau$ will appear initially in the basin around
$\tau_{\rm min}$. Now suppose $y < 6$ when the potential forms. If we
assume that when this happens $\tau$ will slowly roll out towards weak
coupling, then $\tau$ will fall into the minimum when $|x| {\
\lower-1.2pt\vbox{\hbox{\rlap{$>$}\lower5pt\vbox{\hbox{$\sim$}}}}\ }
.42$. If it then stays in the minimum, the total probability that
$\tau$ will settle into the minimum is approximately 12\%. Again this
is not unity, but is not small. Thus no fine tuning is required in the
two condensate model either.

The above discussion is heuristic and preliminary. The dynamics of
$\tau$ (and of other moduli) are under active investigation and will
be described in a future publication~\ref\futpub{Work in progress with
T.~Banks and S.~Shenker.}.

\subsec{Strong coupling pathologies}

$S$-duality provides a natural cutoff at strong coupling.
Ref.~\brustein\ raises the problem that most potentials normally
considered in supersymmetry breaking contain regions where $V(\tau)$
becomes rather large and negative. In general the behavior of $V$ for
$\Im\tau\to 0$ depends sensitively on the values of all the other
moduli (and their potentials). One typically finds $V\to +\infty$ or
$V\to -\infty$ for the general potentials studied in \carlos\ and
\brustein. For example, if $W(\tau,\rho)= \Omega(\tau)/\eta(\rho)^6$
where $\rho$ is a breathing mode, then as $y\to 0$ we have
\eqna\pminf
$$\eqalignno{
V\to +\infty \quad {\rm if} & \quad 3 \left| 1+ 4 i \Im\rho
\p_\rho(\log\eta(\rho))\right|^2>2&\pminf{a}\cr
V\to -\infty \quad {\rm if} & \quad 3 \left| 1+ 4 i \Im\rho
\p_\rho(\log\eta(\rho))\right|^2<2&\pminf{b}\cr
}$$
(assuming $\p_\tau\Omega(\tau)$ is regular at $y\to 0$).
A strong coupling cutoff eliminates the pathology of \pminf{b}.

\subsec{Validity of perturbative/instanton expansions}

Racetrack models with hidden matter typically require large
coefficients of the competing powers of $q$. (For example, in \carlos\
one encounters relative coefficients like $\beta\sim 10^{30}$.) This
raises the problem that higher powers of $q$ might come with higher
coefficients so that it is not justified to keep the two leading
powers of $q$ in a racetrack-like potential. One needs a
nonperturbative calculation, or a nonperturbative principle like
$S$-duality to fix the entire $q$ expansion. Only with such a
principle should one contemplate such large coefficients (if then).

\newsec{Conclusion}

In this note we have focused attention on one example of a moduli
space of string vacua of the general form \emodspcs. It is natural to
ask what will happen with moduli of Calabi-Yau compactification.
Although the analysis will be much more complicated, the essential
properties of $\CM$ are also present in the CY example.  The volume of
Calabi-Yau moduli spaces in the Weil-Petersson-Zamolodchikov (WPZ)
metric is finite~\ref\atdrv{A.N. Todorov, private communication.}.
Moreover, A.~Todorov has proved that the WPZ metric has non-positive
sectional curvature~\ref\todorov{A.N. Todorov, ``The Weil-Petersson
Geometry of the Moduli Space of $SU(n\geq 3)$ (Calabi-Yau) Manifolds
I,'' {\it Commun. Math. Phys.} {\bf 126} (1989) 325.}.  Since compact
manifolds with negative sectional curvature exhibit chaotic motion
under geodesic flow we may expect that generic motions on CY moduli
space will be chaotic. Given the above two facts we can expect that in
weak coupling regions the qualitative aspects of the motion of moduli
described in this paper will survive.

Clearly our discussion has not probed the motion of moduli and the
cosmological consequences of such motion in any great detail. We
intend to return to a deeper investigation of the physics of moving
moduli in a future publication ~\futpub.

\bigskip

\centerline{\bf Acknowledgements}

We are especially grateful to T.~Banks, N.~Seiberg, S.~Selipsky, and
S.~Shenker for many discussions. We thank T.~Banks and N.~Seiberg
for important remarks on the manuscript, and we thank L.~Dixon, D.~Kaplan,
J.~Louis, and A.~Todorov for helpful correspondence. We also thank
V.~Arnold and M.~Gutzwiller for useful remarks on chaotic motion. GM
would like to thank the Rutgers Physics Dept.~for hospitality. This
work is supported by DOE grants DE-AC02-76ER03075, DE-FG02-92ER25121,
DE-FG05-90ER40559, and by a Presidential Young Investigator Award.

\appendix{A}{Some facts about modular functions/forms}

We need a number of facts about modular forms (reviews of modular
functions for physicists can be found in \lerche). First, define
\eqn\eqdef{
q = e^{2 \pi i \tau} \;,
}
and the Dedekind $\eta$-function
\eqn\eeta{
\eta(q) = q^{1/24} \prod_{n = 1}^{\infty}{ (1 - q^n)} \; .
}
We will need the Eisenstein functions
\eqn\eetwo{
E_2(q) = 1 - 24 \sum_{n = 1}^{\infty} {n q^n \over 1 - q^n}
= 1 - 24 \sum_{n = 1}^{\infty}{ \sigma_1(n) q^n } \; ,
}
\eqn\eefour{
E_4(q) = 1 + 240 \sum_{n = 1}^{\infty} {n^3 q^n \over 1 - q^n}
= 1 + 240 \sum_{n = 1}^{\infty}{ \sigma_3(n) q^n } \; ,
}
and
\eqn\eesix{
E_6(q) = 1 - 504 \sum_{n = 1}^{\infty} {n^5 q^n \over 1 - q^n}
= 1 - 504 \sum_{n = 1}^{\infty}{ \sigma_5(n) q^n } \; ,
}
where
\eqn\esigma{
\sigma_k(n) = \sum_{d | n} d^k
}
is the sum of the $k$th powers of the divisors of $n$. We should note
that $E_4(q)$ has a first order zero at $\tau = \rho \equiv e^{ i \pi/3} \sim
e^{ 2 i \pi/3}$, and
$E_6(q)$ has a first order zero at $\tau = i$.

With these definitions in hand, we can write $J(q)$ explicitly as
\eqn\ejdef{
J(q) = { E_4(q)^3 \over \eta(q)^{24} }
= { E_6(q)^2 \over \eta(q)^{24}} + 1728 \; .
}
$J$ is the generator of modular forms of weight zero, and is a
one-to-one mapping from the fundamental region $\CF$ into $\IC\cup
\{\infty\}$. It has a third order zero at $\tau =\rho$,
and $J(q) - 1728$ has a second order zero at $\tau = i$. For small
$q$ (large $\Im \tau$ or weak coupling), $J(q)$ has a Laurent
expansion
\eqn\ejexp{
J(q) = {1 \over q} + 744 + 196884 \> q + \ldots
}
so $J(q)$ has a simple pole at $\tau = i \infty$.

\listrefs

\bye